\documentclass[12pt]{IEEEtran}
\usepackage[cmintegrals]{newtxmath}
\usepackage{cite}
\usepackage{wrapfig, graphicx}
\usepackage{enumitem}

\title{Making Intelligent Reflecting Surfaces More Intelligent: A Roadmap Through Reservoir Computing}
\begin{document}
\author{\IEEEauthorblockN{Zhou Zhou, Kangjun Bai, Nima Mohammadi,  Yang Yi, and Lingjia Liu}
\thanks{The authors are with the Bradley Department of Electrical and Computer Engineering, Virginia Tech, Blacksburg, VA, USA. The corresponding author is L. Liu (ljliu@ieee.org).}}
\maketitle
\begin{abstract}
This article introduces a neural network-based signal processing framework for intelligent reflecting surface (IRS) aided wireless communications systems. By modeling radio-frequency (RF) impairments inside the ``meta-atoms'' of IRS (including nonlinearity and memory effects), we present an approach that generalizes the entire IRS-aided system as a reservoir computing (RC) system, an efficient recurrent neural network (RNN) operating in a state near the ``edge of chaos''. 
This framework enables us to take advantage of the nonlinearity of this ``fabricated'' wireless environment to overcome link degradation due to model mismatch. 
Accordingly, the randomness of the wireless channel and RF imperfections are naturally embedded into the RC framework, enabling the internal RC dynamics lying on the edge of chaos. Furthermore, several practical issues, such as channel state information acquisition, passive beamforming design, and physical layer reference signal design, are discussed. 
\end{abstract}


\section{Introduction}
With the unfolding of the 5th generation cellular systems (5G), beyond 5G (B5G) and 6G technologies are rapidly becoming new buzzwords in telecommunication academia and industry. The wireless market analyses suggest that we are expected to witness the demand for trillions of wireless connections and equipment energized by the surge of affordable and low-energy devices \cite{shafin2020artificial}. This unprecedented number, however, poses complications and challenges to the cellular communications systems design. Intelligent reflecting surface (IRS) assisted wireless networks are deemed as promising solutions 
\cite{bjornson2020reconfigurable,bjornson2019intelligent, bjornson2019massive,wu2020intelligent}, where IRSs are utilized to configure a more ``favorable'' wireless channel between access points (APs) and mobile stations (MSs). IRS controls each reflecting unit (``meta-atoms'') towards an ideal direction to form a fine-grained reflecting beam. 
Furthermore, IRS is primarily implemented by passive RF components, reducing the cost compared to massive MIMO/small-cell/relay aided systems that resort to active antenna units. Thereby, by adding these configurable IRS components to the wireless environment, an extra degree of freedom (DoF) in the wireless environment can be achieved \cite{bjornson2019intelligent}. By properly adjusting the reflecting components on the IRS, the newly formed beam can constructively enhance the desired signal or destructively anneal interference power amid transmission streams.
Overall, the introduction of IRS forges a new path for the realization of a programmable and intelligent channel environment, increasing the network spectrum and energy efficiencies \cite{wu2020intelligent}. 

\subsection{Main Challenges}
Although IRS is deemed as a cost-efficient approach toward high system efficiency, continuing innovations on hardware and software of IRS are yet imperative for the realization, where the following critical issues are emerged to be addressed:
\begin{itemize}[noitemsep, nolistsep, leftmargin=1.0pc]

\item \textbf{Hardware Impairments:}
Current methods and apparatus devised for IRS-aided cellular systems have been developed with the premise of ideal radio-frequency components as well as a simplistic mathematical modeling approach, such as the element-wise phase-shift model \cite{wu2020intelligent}. However, the presence of memory effects and nonlinearity in hardware, particularly when the operation is configured on the entire frequency band, may lead to model mismatch which can deteriorate the performance compared to the theoretical setup.
\item \textbf{Reflection Optimization:}
An ideal IRS-aided operation is achieved through jointly optimizing the reflection phase on IRS, a.k.a. passive beamforming, as well as the active beamforming/receiving on APs and MSs. However, the inherent practical constraints, such as the finite alphabet feature of the tunable phase dictionary, the duplex mode, and the asymmetric number of antennas configured on MSs and APs, make the problem vastly complicated. Furthermore, hardware implementations on the meta-atoms, either electronically or mechanically, have shown potential bottlenecks on achieving adequately fast phase shifting, in which the passive beamforming may lag behind the variations on environment and user mobility. Needless to say, the problem space would render a na\"ive exhaustive search intractable. 

\item \textbf{Channel Acquisition:} To accomplish the joint beamforming/receiving design, we require multi-fold channel state information (CSI) which includes links from AP to IRS, from IRS to MSs, from AP to MSs, and all the reversed links. For indirect channels (links with IRS at one end), it is challenging to obtain the CSI because IRS is deployed without transceiver chains, which makes the reference signal on IRS unavailable to access. Meanwhile, IRS is often configured with a large amount of ``meta-surface''\cite{ozdogan2019intelligent}, which makes the channel dimension and the overhead on the CSI estimation prohibitively high. More importantly, the online reference signals in wireless communications systems are very limited due to training overhead constraints defined in communication standards\cite{shafin2020artificial}. Besides, environment scatters are located within the Rayleigh distance of the IRS, which can fundamentally change the channel model due to spherical wavefront features. 
\end{itemize}

\subsection{State-of-the-Art}
There are attempts in the literature to address the issues raised above:
\subsubsection{Hardware Impairments}
Recent work \cite{hu2018capacity} analyzes the channel capacity loss by hardware impairments. The analysis characterizes the imperfectness of the meta-atoms on IRS as additive Gaussian noise. It shows that both the system capacity and the derivative of the capacity with respect to the area of meta-surface decrease when the scale of reflective components increases. The research in \cite{badiu2019communication} has investigated that when the hardware impairments are modeled as phase errors, the transmission can be equivalently treated as a point-to-point Nakagami fading channel. While these are seemingly reasonable analytical results, they do not provide feasible solutions to the hardware impairments. 

\subsubsection{Reflection Optimization}
Rather than using active RF components to harness spatial diversities (such as relay, and backscatter communications, etc.), IRS seeks to optimize spectral efficiency and at the same time to consume less energy. To address the aforementioned challenges raised by the discrete phase values and large meta-atoms, \cite{wu2019intelligent} introduced a heuristic alternating optimization technique that relaxes the discrete variables as continuous variables. Similarly, energy efficiency in terms of bit-per-Joule can be considered in the same optimization framework. This method is infeasible in real-time and does not explore the dynamics of user mobilities and channel environments. Another related work also extended the optimization on the joint design of pilot patterns and network utilities. Nonetheless, they typically rely on unrealistic assumptions on perfect channel knowledge, synchronous controls, and unlimited user feedback. 

\subsubsection{Channel Estimation}
Explicit channel state information is considered an essential component of the reflection optimization in IRS systems. Due to limited RF chains on the reflective surface, the uplink and downlink reciprocity has been widely leveraged in the channel estimation. In \cite{taha2019enabling}, it introduced using compressed sensing and deep learning-based approaches to infer the entire channel by using only a relatively small number of accessible active RF components on the IRS, where the sparsity/low-rank properties of the channel are leveraged. However, when receiving RF chains are not installed at the IRS, this method becomes infeasible. A viable way to solve this challenge is to use implicit channel information instead. \cite{wu2019towards} introduced learning a beamforming vector via adding a feedback channel in the IRS system. However, the feedback overhead becomes prohibitively high as the number of meta-atoms increases.

\subsection{Our Contributions}

The contribution of this paper is the following: 
\begin{itemize}[noitemsep, nolistsep, leftmargin=1.0pc]
\item Unlike the existing formulation of IRS, we consider modeling both the nonlinearity and memory effects of each IRS component, mimicking practical RF circuits behave, where the memory effects are generally not identified and widely overlooked in the current literature.
\item By leveraging the similarity between hardware uncertainties of IRS systems and chaotic features of reservoir computing (RC) systems, we generalize an entire IRS-aided system as an RC. 
This approach does not rely on explicitly modeling the hardware impairments, whereas it could circumvent the model mismatch by universal approximation properties of the RC. 
This design offers a promising direction on bridging artificial intelligence (AI) and wireless systems.
\item We point out some of practical issues in the implementation of this RC based framework and further investigate possible solutions such as reference signal design and training algorithms.
\end{itemize}
The remainder of this paper is organized as follows. In Sec. \ref{Sys}, we review the formulation and hardware impairments of IRS systems. In Sec. \ref{RC}, we introduce the RC framework and its mapping to IRS. The practical issues of applying this technique are discussed in Sec. \ref{discuss}. Finally, the paper is concluded in Sec. \ref{concl}.

\section{System Formulation and Hardware Model}
\label{Sys}
\begin{figure}[!t]
\includegraphics[width=0.9\linewidth,trim={1.8cm 6.5cm 3cm 7.2cm},clip]{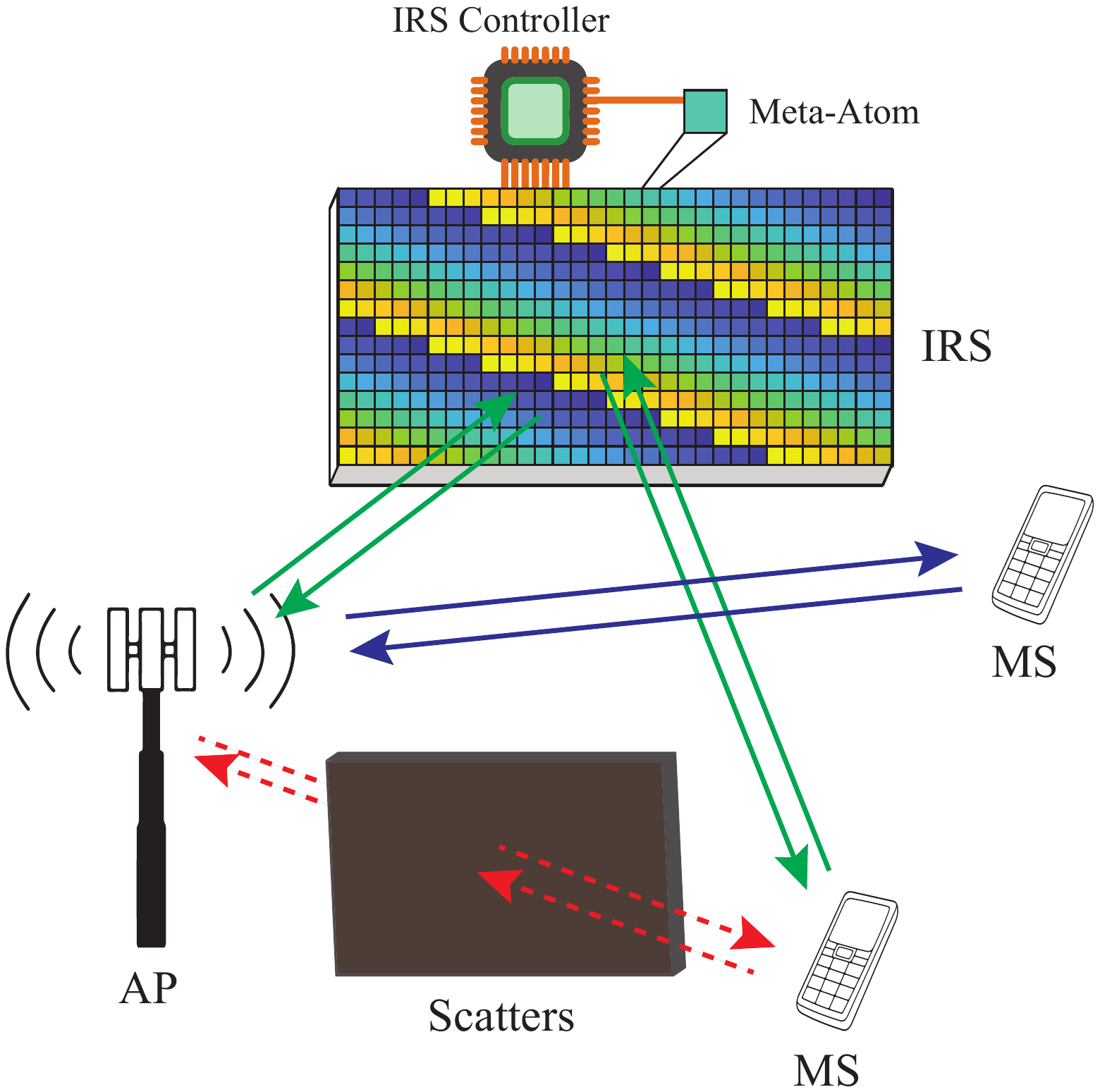}
\caption{An illustration of an IRS-aided wireless system: multiple MSs, one IRS, and one AP.}
\label{fig_scenario}
\vspace{-7mm}
\end{figure}

We first consider an ideal model of IRS-aided wireless communications systems where multiple mobile stations (MSs) simultaneously communicate with one access point (AP) associated with a single IRS as depicted in Fig. \ref{fig_scenario}. The IRS is equipped with passive reflecting units, namely, "meta-atoms". The entire wireless transmission channel can be conceptually partitioned into three parts (in uplink): a forward channel from MSs to IRS, a reflecting channel from IRS to AP, and a direct channel from MSs to AP. A zero gain of the direct channel indicates a transmission blockage. The downlink channel also follows in a similar naming approach. 

\subsection{Conventional IRS Formulation}

According to the standard definition of IRS, the incoming electromagnetic waves are reflected in the desired direction by using predefined phase-shifts. The conventional embodiment of IRS is constructed by massive meta-atoms, each of which is enabled with software-controllable properties on configuring the reflecting direction. The entire formed meta-surface passively forwards RF signals with modified amplitude and phase, in which the operation is fundamentally different from the concept of amplify-and-forward relay which requires active RF chains \cite{bjornson2019intelligent}. The reconfigurability is either achieved electronically, by means of positive-intrinsic negative (PIN) diodes or field-effect transistors (FETs), or mechanically via microelectromechanical system (MEMS) switches.

Conventionally, each meta-atom is assumed to be capable of changing the amplitude and phase of the incident signal and constructing a reflection towards a new direction independently. In other words, no signal-coupling effects between IRS units are assumed. A math formulation of this perfect reflection is via a diagonal matrix. Each diagonal entry represents the imposed amplitude and phase change to the incident signal contributed by the corresponding meta-atom. Optimizing these entries can bring an improved `augmented' channel between MSs and AP. However, this formulation, governed by the ideal physical characteristics of the meta-surface can lead to model misspecification issues as we pointed out in the introduction.

\subsection{Hardware Nonlinearity in IRS}
\begin{figure}[!t]
\includegraphics[width=0.9\linewidth,trim={4cm 11.7cm 4.6cm 10cm},clip]{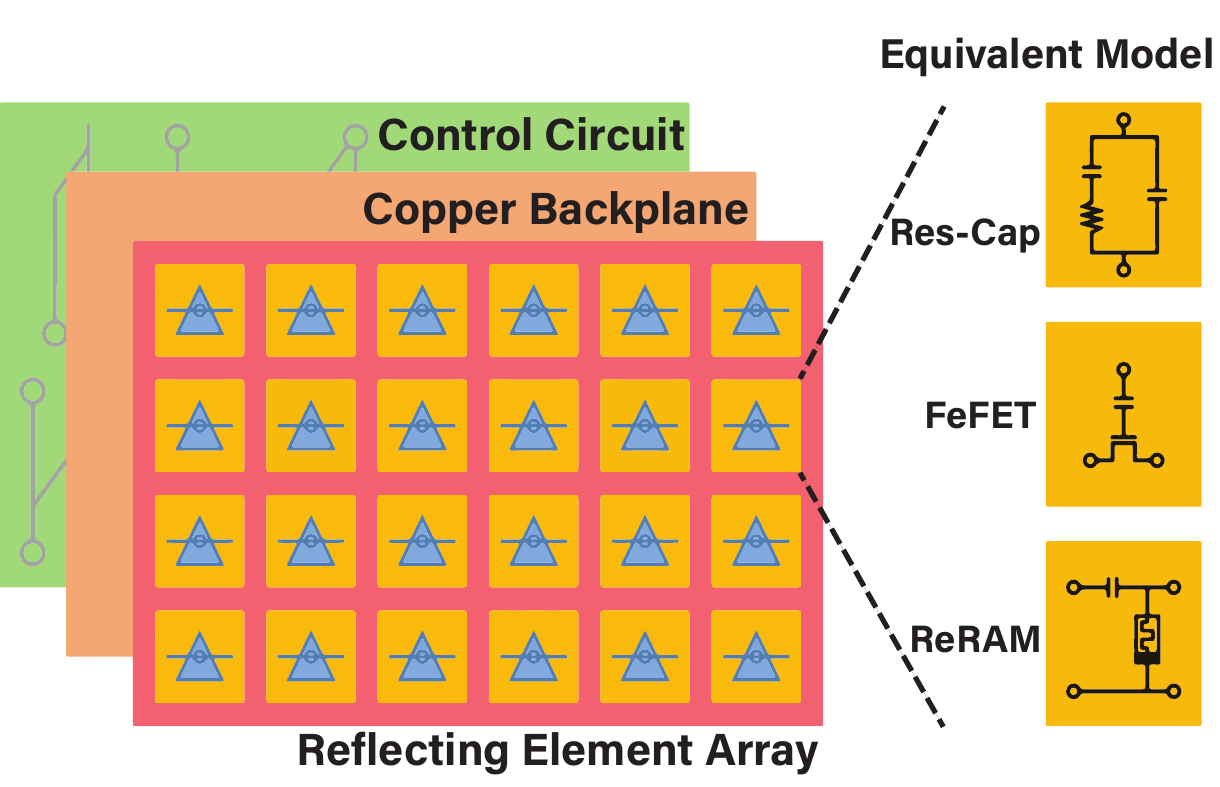}
\caption{A realization of an IRS made of a control circuit, a copper plate and an array of reconfigurable reflecting elements.}
\label{fig:irs-hw}
\vspace{-7mm}
\end{figure}

Overall, an IRS is composed of layered materials, generally three layers, as demonstrated in Fig. \ref{fig:irs-hw}. The outer layer is made of an array of reflecting elements printed on a dielectric substrate, in which the reflecting elements can interact with RF signals. This would provide desired amplitude and phase shifts to incident signals based on the device characteristics, including the geometrical alignment between reflecting elements. The intermediate layer is built of a copper plate to avoid energy leakage from incident signals. Lastly, the inner layer is a control circuit to calibrate reflecting elements in real-time based on the desired amplitude and phase. An ideal IRS implementation is considered as with this reconfigurability without nonlinearity and memory effects. However, in practice, the memory effect is inherently subsistent to devices due to the charge accumulation along with the reactive-ion etching during the fabrication process.

Passive devices could satisfy the necessity of reconfigurability as required by the IRS. The second-order resistor and capacitor (Res-Cap) model is a simple implementation approach where a digitally-controlled resistor with two adjacent capacitors are employed. By controlling the model's resistance state, the resistor can be switched between its high-resistance-state (HRS) and Low-resistance-state (LRS). Eventually, energy from RF signals is dissipated according to the resistance state, and thus, controlling the reflection amplitude. 
Furthermore, various phase shifts of the reflecting elements can be realized independently as different time constants are achieved based on the resistance state.

In recent years, the Ferroelectric Field-Effect Transistor (FeFET) \cite{si2019ferroelectric} and the Resistive Random-Access Memory (ReRAM) \cite{an2020robust} have been introduced in neuromorphic applications as two emerging building blocks for emulating artificial neural networks. This makes these emerging devices possible candidates for implementation of IRS, due to their intrinsic reconfigurability.
The FeFET, a descendant of the MOSFET, consists of a ferroelectric layer in between its electrode and conduction region. The ferroelectric thin film could remember the electric field to which it had been exposed, switching the capacitance of the device from high to low, or vice versa. By controlling the biasing signal across the device, different amplitude and phase changes can be achieved as the energy dissipated in the device is controllable. Similar to the classical Res-Cap model, the ReRAM is a controllable resistor through its conductive filaments. By altering the biasing signal across the device, various resistance states can be achieved, and thus, realizing different amplitude and phase changes.


In practice, the ideal linearity assumption of reflecting units cannot be satisfied due to the inherent properties thereof. We can extend our previous formulation by including a time-dependent state equation, where the current state would be determined by the previous state as well as the current incident signal, and the state transitions are according to a nonlinear function. This new formulation leads to a modeling scheme that accounts for both nonlinearity and the memory effects. 

\section{Reservoir Computing-Based Reformulation}
\label{RC}
In this section, we begin by briefly introducing the architecture of RC. Then, we discuss how RC is analogous to IRS systems. Finally, we elaborate on how the main challenges (pointed out in the introduction) are accordingly addressed.  

\subsection{Preliminary}
\begin{figure}
\centering
\includegraphics[width=1\linewidth, height = 0.5\linewidth]{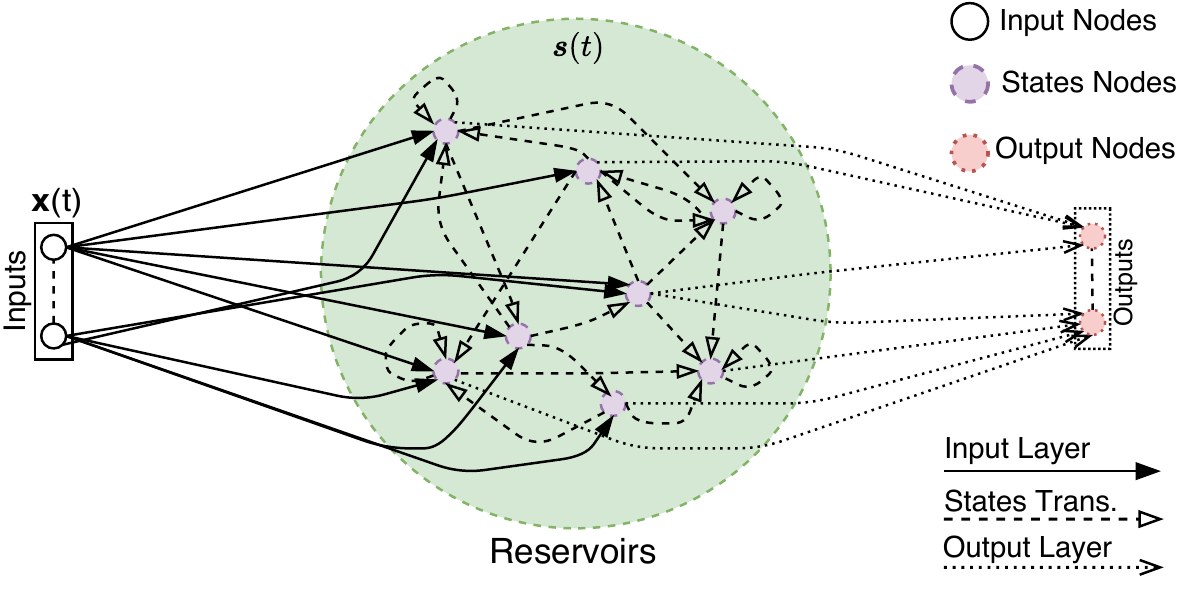}
\caption{A realization of reservoir computing - echo state network: the input is first mapped to a high dimensional space to create reservoir states, then the output is obtained through a readout mapping.}
\label{fig:pic-2}
\vspace{-6mm}
\end{figure}
RC is a computational framework motivated by RNN mechanism which consists of three parts: an input mapping, a fixed dynamic system, and a trained readout network. A realization of RC, echo state network (ESN)\cite{zhou2020deep}, is illustrated in Fig.~\ref{fig:pic-2}. The constitutive modules of an ESN are as follows:
\subsubsection{\textbf{Input Mapping}}
The ESN input is first projected to a higher dimensional space by multiplying to a weight matrix that is initialized with random values. Mapping the input signal into a high-dimensional space is useful for pattern analysis which is a shared concept of kernel-based methods. 

\subsubsection{\textbf{Reservoir Dynamics}}
The underlying reservoir dynamics are characterized by a recurrent equation. The equation basically describes a Markov process of the internal states stimulated by the input. In addition, an activation function is added between two consecutive internal states to offer nonlinearity. 
Furthermore, the internal state transition matrix stays fixed during the training stage. In practice, improperly initialized internal weights often lead to diminishing generalization performance. To circumvent this issue, certain algebraic properties of the transition matrix has to been satisfied, such as echo state property (ESP) which states that given an input sequence, the output should not differ far from two different initial states, that is, the effect of initial conditions should vanish as time proceeds. 

\subsubsection{\textbf{Learning Readout}}
Given the states of reservoir, the output is obtained via a dimension reduction process on a subset of the reservoir states. The coefficients of the readout mapping are determined via minimizing a predefined loss function, where the loss measures the dissimilarity between targets and predicted outputs. The optimization on the loss can be either solved using gradient descent or through the least-squared framework when Frobenius norm is selected as the loss function.

\subsection{Reformulation}
\looseness=-1
Now, we consider how the IRS-aided wireless system can be treated as an RC system, specifically as an end-to-end RC-based auto-encoder. We analogize the signals sent out from all active stations (APs or MSs) to the IRS as the input of the RC; thereby, the IRS serves as the reservoir. Accordingly, the readout layer is considered as a three-layered feed-forward neural network with identity activation functions, which is comprised by IRS coefficients, reflection channel, and receiving processing matrices for the uplink or precoding matrix for the downlink at AP. The link between active stations and IRS is considered as the input layer of the RC. A similar idea of analogizing wireless communication systems to neural networks can be found in \cite{Com_survey}, where IRSs are modeled as feed-forward neural networks. However, the formulation is not sufficient to capture the inherent dynamics (by hardware memory effects) of IRS. Accordingly, we optimize a loss function that measures the link-level transmission reliability or efficiency using the passive beamforming matrix, active receiving and precoding matrices as variables. Technically, the loss function is defined as a distance between the received signal and the desired signal using training dataset. 

\subsection{Remarks}
Regarding the challenges enumerated in the introduction, the introduced RC-based approach has the potential to address them, as elaborated below:
\subsubsection{Hardware Impairments}
The hardware impairments are solved since the nonlinearity and memory effects are circumvented by altering the distortion as needed nonlinear internal state transitions of RC, where chaotic internal states can create rich representations of the input. On the other hand, deploying more meta-atoms on the IRS is equivalent to using higher-dimensional internal states which can reduce the learning loss value, however, increases the risks on overfitting.
\subsubsection{Passive Beamforming}
The optimization procedure on the phase-shift and amplitude adjustment is naturally considered as learning one of the output layers of RC. Meanwhile, the receiving processing matrix is considered as the final output layer of the RC, whereas the active precoding matrix is treated as the first layer of the RC in downlink transmission. Since the memory ability is enabled by RC, the learning can essentially capture the underlying long term features of the channel environment. Thus, the learned beamformer is to act in a proactive manner to handle the interference by leveraging the historical data, which can potentially solve the lag effects brought by the computation. 

\subsubsection{Channel Acquisition}
\looseness=-1
In the joint optimization on the uplink passive and active beamforming, the CSI from MSs to IRS is no longer needed. This is because it is treated as the input layer of RC, which does not require to be known in the framework of RC. Conversely, the downlink active precoding and passive beamforming can be obtained by using channel reciprocity. More importantly, the channel environment between AP and IRS is relatively static, where this static property can be further leveraged to reduce the training overhead. Therefore, the reference signal resources for CSI acquisition would be significantly reduced compared to conventional approaches. 

\begin{figure}
    \centering
    \includegraphics[width=0.9\linewidth]{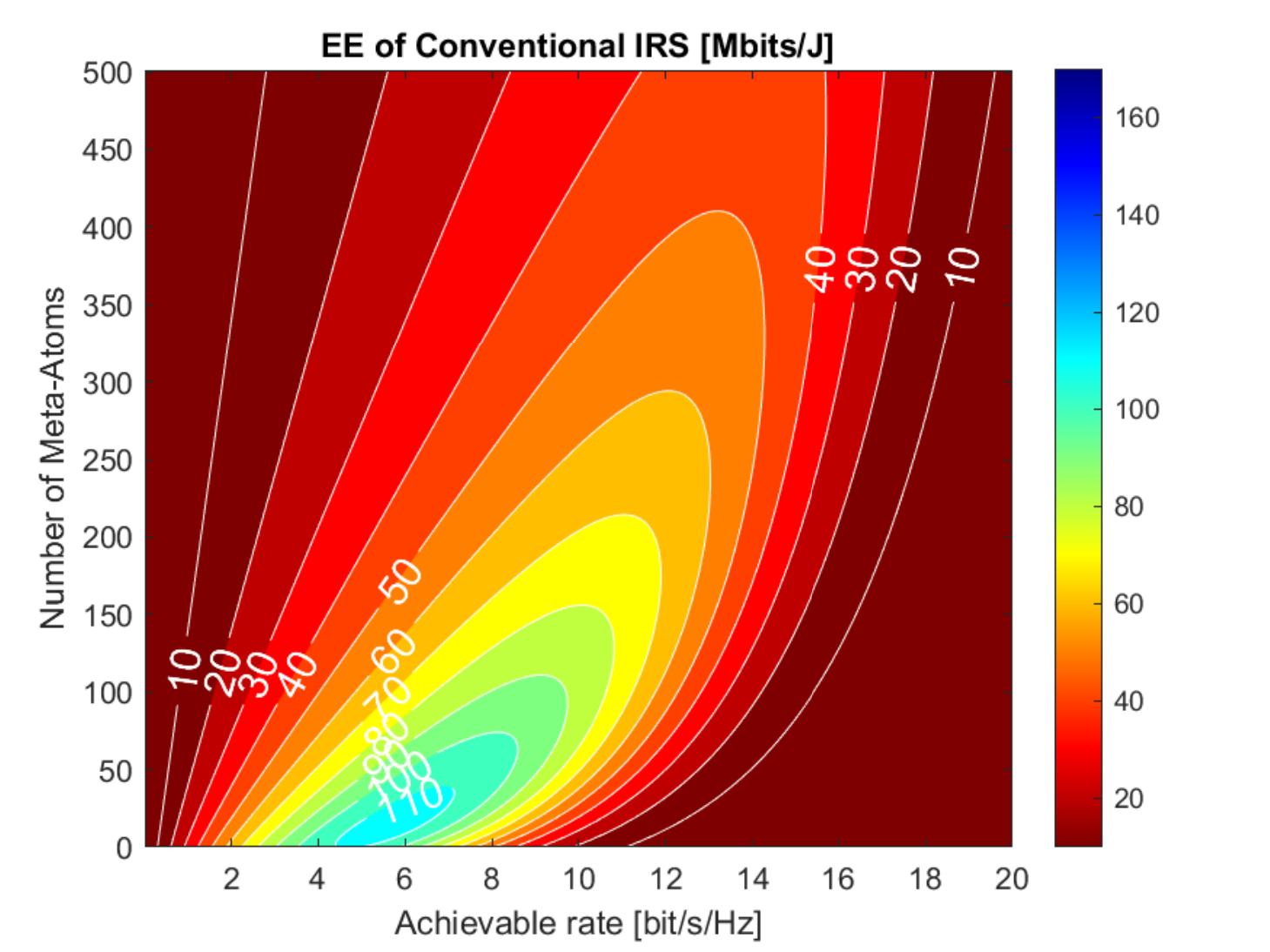}
    \hspace{0mm}
    \includegraphics[width=0.9\linewidth]{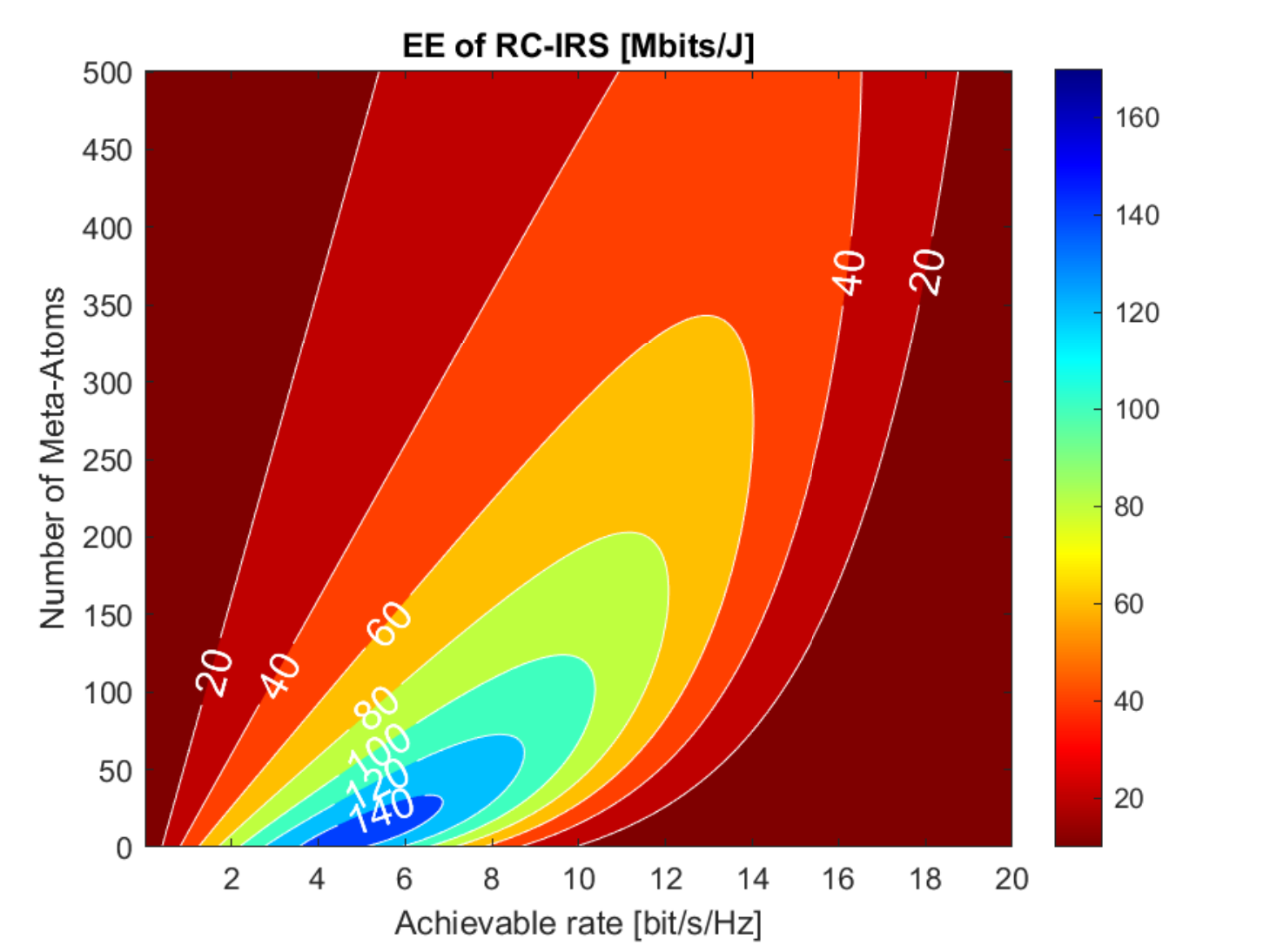}
    \caption{Energy Efficiency evaluation of (a) a model-based optimization approach, and (b) RC based learning approach in an IRS system which is properly trained.}
    \vspace{-5mm}
    \label{fig:theoretical_comparison}
\end{figure}

\looseness=-1
A comparison between RC-based IRS and the method introduced in \cite{bjornson2019intelligent} (referred as a model-based approach) is presented in Fig. \ref{fig:theoretical_comparison}. For simplicity, the evaluation scenario is configured with one MS, one IRS and one AP. In our analysis, the hardware impairments are equivalently characterized as channel estimation errors and power dissipation at IRS, where the CSI is utilized by model-based approach for passive beamforming design, and power dissipation leads to additive interference on the transmission link. Remark that more meta-atoms can lead to stronger interference though it potentially offers higher beamforming gain. The RC-based approach demonstrates advantages over the conventional one in terms of energy efficiency (EE), which can be attributed to its ability to overcome model mismatch by leveraging training dataset.

\section{Discussion}
\label{discuss}
The RC framework is envisioned as an efficient (in terms of spectrum and energy) and effective (in terms of handling hardware impairments) signal processing framework for passive and active beamforming/receiving design of the IRS system thanks to the chaotic and high dimensional features of the reservoir. In this section, we discuss related practical issues of using this framework. There are two prerequisites for conducting the learning for the RC: 1) the reference signals for adjusting link-level demodulation, namely, training dataset; and 2) the CSI from IRS to AP and from MSs to AP for formulating the learning loss function. To this end, we can directly leverage the reference signal structure defined in 5G New Radio (NR) standards. For instance, the reference signals are comprised of demodulation reference signals (DMRS) and the channel state information reference signals (CSRS). Accordingly, DMRS can be utilized as the training dataset, whereas CSRS can be applied to track the CSI. Since the channel environment from IRS to AP is relatively static, the CSI of IRS-AP can be initialized by a channel sounding process at the very beginning of the IRS deployment.

In an attempt to track environmental changes between IRS and AP, we introduce a channel acquisition method based on the uplink-downlink reciprocity. 
Remark that only AP and MS can send the reference signals due to the availability of active RF chains. To facilitate the estimation, we assume that full-duplex transmission is enabled at AP. In this method, AP sends CSRS to IRS before receiving the reference signals reflected back by the IRS. Since the received reference signals contain the back and forth channel information between AP and IRS, the channel variation can be accordingly tracked by learning common features of the row space and column space of the received reference signals. 
In addition, we can incorporate a round-robin scheme into this method by estimating the channel variations of a subset of the IRS units at each turn. Accordingly, the entire CSI of the AP-IRS is attained after all the meta-atoms are traversed.



Furthermore, the RC-based framework can be extended to other chaotic neural networks based learning paradigms, such as extreme learning machines. However, structurally and operationally, there is a clear agreement between IRS and RC-based networks that is absent for other neural networks such as LSTM. In addition, LSTM is very challenging to train, whereas RC is known to achieve better generalization performance compared to other NNs using extremely limited training dataset \cite{zhou2020deep}, where the limitation on the training data is one of the fundamental challenges raised in the introduction. The RC characterized IRS also can be incorporated as a policy network in a reinforcement learning (RL) framework. Accordingly, the objective becomes maximizing the cumulative reward of the underlying policy with passive and active beamforming as the actions. In general, this RL enabled approach is more robust in dynamic settings - where it operates the passive and active beamforming in a proactive way.


\section{Conclusion}
\label{concl}
In this article, we provided a roadmap for the application of RC-based processing techniques in the design of IRS-aided wireless systems.
We drew parallels between reservoir computing and the IRS transmission link, and introduced an ideal mapping thereof. 
We also pointed out an equivalence between learning readout layers of RC and designing passive beamforming as well as active receiving. 

\bibliography{IEEEabrv,./ref}

\end{document}